\documentclass[12pt]{article}
\usepackage{amssymb,amsmath,amsfonts,mathrsfs}
 
\usepackage{graphicx}
\usepackage{geometry}
\usepackage{hyperref}

\providecommand{\keywords}[1]
{
  \small	
  \textbf{\textit{Keywords---}} #1
}
\title{Quantum Analysis of BTZ Black Hole Formation Due to the Collapse of a Dust Shell}
\author{Alexander A. Andrianov$^{1,2,\footnote{sashaandrianov@gmail.com}}$, Artem Starodubtsev$^{1,\footnote{artemstarodubtsev@gmail.com}}$ and  Yasser Elmahalawy$^{1,3,\footnote{yasserreda99@gmail.com}}$ \\
        \small $^{1}$Department of High Energy and Elementary Particles Physics, Saint Petersburg State \\ \small  University, Saint Petersburg 198504, Russia;  \\
        \small $^{2}$Departament de F\'{ı}sica Qu\`{a}ntica i Astrof\'{ı}sica and Institut de Ci\`{e}ncies del Cosmos (ICCUB), \\ \small Universitat de Barcelona, Mart\'{ı} i Franqu\`{e}s 1, 08028 Barcelona, Spain; \\
        \small $^{3}$Department of Physics, Faculty of Science, Benha University, Banha 13518, Egypt. \\
}
\date{} 

\begin{document}
\maketitle
\begin{abstract}
We perform Hamiltonian reduction of a model in which 2+1 dimensional gravity with negative cosmological constant is coupled to a cylindrically symmetric dust shell. The resulting action contains only a finite number of degrees of freedom. The phase space consists of two copies of $ADS^2$ -- both coordinate and momentum space are curved. Different regions in the Penrose diagram can be identified with different patches of $ADS^2$ momentum space. Quantization in the momentum representation becomes particularly simple in the vicinity of the horizon, where one can neglect momentum non-commutativity. In this region, we calculate the spectrum of the shell radius. This spectrum turns out to be continuous outside the horizon and becomes discrete inside the horizon with eigenvalue spacing proportional to the square root of the black hole mass. We also calculate numerically quantum transition amplitudes between different regions of the Penrose diagram in the vicinity of the horizon. This calculation shows a possibility of quantum tunneling of the shell into classically forbidden regions of the Penrose diagram, although with an exponentially damped rate away from the horizon.
\end{abstract}


\section{Introduction}
 General Relativity encounters problems at short distances both at classical and quantum level. Classical gravity develops singularities, while there are ultraviolet divergences in quantum gravity which cannot be removed by renormalization.
 
On the other hand, as it was first argued by Bronstein \cite{mpb} there is the smallest possible distance in quantum gravity, the Planck length, beyond which measurements are not possible. The argument relies on non-perturbative effects such as black hole formation and could be described only within a non-perturbative quantum theory of gravity absent to the date.

In the absence of a full theory, non-perturbative quantization can be performed for some symmetry-reduced models for General Relativity in which all but a few degrees of freedom are removed \cite{lqc}. Within such models, black hole formation could be described, which is essential for Bronstein's argument.

The simplest model of this kind is gravity coupled to a spherically symmetric thin dust shell. This model was extensively studied both on classical \cite{an,israel1,kuchar,hajicek} and quantum \cite{louko,vb,vb1,hk} level. In some of this work, a resolution of singularity was obtained  \cite{vb,hk}. However, the quantum theories obtained in different work are not equivalent. This may be a result of quantization ambiguity as well as the non-trivial structure of the phase space of the model. The definition of a wavefunction on different sectors of the configuration space of the model can lead to inequivalent theories. 

In quantum theory, it is common that the definition of the wavefunction has to be extended to all possible configurations, whether they are classically reachable or not. In a particular way, it was realized in \cite{vb,vb1} where the phase space was complexified and its different sectors were assembled into a Riemann surface where the branching point represented the horizon.

On the other hand, there is an example where the phase space of a similar model was given a real global chart. This model is gravity in 2+1 spacetime dimensions coupled to a point particle \cite{thooft,thooft1,mw}. The momentum of the particle turns out to be an element of the  Lorentz group, and the Hamiltonian constraint fixes the conjugacy class of this group element.

An attempt to relate the two above approaches was made in our previous work \cite{aes1,aes2} for zero cosmological constant. The momenta turned out to form $ADS^2$ space, which results, in particular, in the non-commutativity of the coordinates. The Hamiltonian constraint was found to be different from that of a particle, accounting for the gravitational field generated by inter-shell movement energy. The relation between group-valued momenta and canonical momenta analogous to that of \cite{vb,hajicek} was found. Transition amplitudes between zero and positive shell radii were found to show no divergences, which could be interpreted as singularity resolution in quantum theory.

However, the above model is substantially different from 3+1 dimensional gravity as it has naked singularity solutions instead of black hole solutions. The closest analog to 3+1 dimensional model is 2+1 dimensional gravity with negative cosmological constant which has the well-known BTZ black hole solutions \cite{BTZ}.

BTZ black holes has been extensively studied both semiclassically \cite{marc17,marc19}, 
where the back reaction of a quantum field on spacetime geometry was taken into account, and via holographic approach \cite{carlip,re}, using boundary conformal field theory. Here we will consider a simpler model with axial symmetry, because this model could be generalized to higher spacetime dimensions. 

In this paper, we partially extend the results of \cite{aes1,aes2} to the case of negative cosmological constant. We focus on studying quantum dynamics in the near-horizon area, where quantization could be performed by traditional methods due to momentum commutativity. The question of interest to be asked in this regime is the possibility of quantum tunneling into classically inaccessible sectors of the Penrose diagram.

The paper is organized as follows. In section \ref{sec2}, we reproduce the results of \cite{vb,hajicek} for 2+1 dimensional gravity+dust shell system with negative cosmological constant. The only difference between this result in 2+1 and 3+1 dimensional gravity is the absence of Newtonian potential and a contribution from  cosmological constant, but the solution to the constraints branches in the same way.

In section \ref{sec4}, the results of \cite{thooft,thooft1} and \cite{mw} are generalized from a point particle to a circular shell, describing the later as an ensemble of circularly arranged point particles by analogy with \cite{aes1,aes2}, but now with negative cosmological constant. As before, the momenta of the shell form $ADS^2$ space, and coordinates are again non-commutative.

In sections \ref{sec5} and \ref{sec6},  we derive the expression for the Hamiltonian constraint of the model in terms of global phase space coordinates. This  constraint turns out to be slightly  different from that of \cite{aes1,aes2} due to partial compensation of the positive curvature created by the shell by negative curvature from the cosmological constant term. The relation between momenta from $ADS^2$ and canonical momenta from \cite{vb,hajicek} is found.

In section \ref{sec7}, we consider an approximation  when the shell is located close to the horizon. This allows us to quantize the model in momentum  representation on $ADS^2$. We find that  shell coordinates are non-commutative and  one of them (time) has a discrete spectrum. The  shell radius is found to have  discrete spectrum inside the horizon and continuous spectrum, but separated from zero, outside the horizon.

In section \ref{sec8}, we study the quantum dynamics of the shell. We derive the expression for the evolution operator and calculate numerically some of its matrix elements. These matrix elements  describe the transition amplitudes between different sectors of the Penrose diagram. It turns out that there is a possibility of quantum tunneling for the shell into classically not accessible regions. Generalization of some of these results to 3+1 dimensional gravity is also discussed.

\section{2+1 gravity coupled to a dust shell and its ADM canonical analysis} \label{sec2}
The general metric of a cylindrically  symmetric spacetime \cite{hajicek,vb} is
\begin{align}
\label{13}
ds^{2} & = -\big( N^{2} - L^{2} (N^{r})^{2} \big) dt^{2} + 2 L^{2} N^{r} dtdr + L^{2}dr^{2} + R^{2} d \theta^{2},
\end{align}
where $N$ and $N^{r}$ are lapse and shift functions. Note that $N$, $N^{r}$, $L$, $R$ are continuous functions describing the gravitational field. The ADM analysis leads to the following canonical action
\begin{align}
\label{12}
S _{\Sigma}[R,L,P_{L},P_{R};N,N^{r}] & = \int dt  \int _{-\infty}^{\infty} dr (P_{L} \dot{L} + P_{R} \dot{R} - NH^{G} - N^{r}H_{r}^{G} ),
\end{align}
where the constraints are
\begin{align}
\label{12a}
H^{G} & = - P_{L}P_{R} + L^{-1} R'' - L^{-2} R' L',
\end{align}
and
\begin{align}
\label{12b}
H_{r}^{G} & = P_{R}R' - L P'_{L}.
\end{align}
It could be simplified by Kucha\v{r} canonical transformation \cite{kuchar} as\footnote{The notation for mass here is different from that of the original paper \cite{BTZ} by substituting $-M \rightarrow 1-2m$, so that the empty ADS space is recovered at $m=0$. Note also that the cosmological constant is $|\Lambda| = 1/l^2$, where $l$ is the radius of curvature.}
\begin{align}
\label{15a}
L & =  \sqrt{ R^{'2} (1-2m+ |\Lambda| R^2)^{-1}- (1-2m+ |\Lambda| R^2) P^2_m},   \\
\label{15b}
P_{L} & =\frac{ (1-2m+ |\Lambda| R^2)P_{m}}{\sqrt{ R^{'2} (1-2m+ |\Lambda| R^2)^{-1} - (1-2m+ |\Lambda| R^2) P^2_m}},   \\
\label{15c}
\bar{R} & =  R,  \\
\label{15d}
\bar{P}_{R} & =  P_R -  \frac{(1-2m+ |\Lambda| R^2)^{-1}}{L^{2}}   \Big[ (L P_{L})' R' - (L P_{L}) R''  \Big],
\end{align}
to give Liouville form
\begin{align}
\label{16}
\Theta & =  \int   P_{L}  \dot{L} + P_{R}  \dot{R},  \\
\label{17}
 & =  \int   P_{m}  \dot{m} + \bar{P}_{R}  \dot{R} + \frac{\partial}{\partial t} \Bigg[ L P_{L} + \frac{R'}{2} \ln  \Bigg| \frac{R' - L P_{L}}{R' + L P_{L}}  \Bigg| \Bigg] +  \frac{\partial}{\partial r} \Bigg[ \frac{ \dot{R}}{2} \ln \Bigg| \frac{R' + L P_{L}}{R' - L P_{L}}  \Bigg| \Bigg],
\end{align}
with a simple set of constraints:
\begin{align}
\label{18}
\bar{P}_{R} & =0, ~~~~~ m'=o.
\end{align}
Now, we turn to gravity+shell system. The total action in this case is
\begin{align}
\label{19}
S & =  S_{gr} + boundary~ terms + S_{shell},   \nonumber  \\ & = \frac{1}{16\pi} \int  (R - 2\Lambda) \sqrt{-g} d^{3}x + (boundary~ terms) + M \int _{\Sigma}  d\tau.
\end{align}
The first  term in (\ref{19}) is  the  Einstein-Hilbert action. The third term is the shell action given by
\begin{align}
\label{20a}
S_{shell} & =  -  M \int _{\Sigma} \sqrt{\hat{N}^{2} - \hat{L}^{2}(\hat{N}^{r}+\dot{\hat{r}})^{2}} dt,
\end{align}
where the fields evaluated at the location of the shell are denoted by hats and $M$ is the bare mass of the shell. The action (\ref{19}) in the Hamiltonian form is 
\begin{align}
\label{20}
S & = \int \hat{\pi} \dot{\hat{r}} dt +  \int  \Big[ P_{L} \dot{L} + P_{R} \dot{R} - N (H^{s} +H ^{G}) - N^{r} (H_{r}^{s} - H_{r}^{G}) \Big] dr dt + \int m_{ADM} dt ,
\end{align}
where $m_{ADM}$ is the total mass of the shell, which takes into account the gravitational mass defect, and $ \hat{\pi}$ is the momentum conjugate to $ \dot{\hat{r}}$, which is equal to
\begin{align}
\label{21}
\hat{\pi} & = \frac{M \hat{L}^{2} (\hat{N}^{r}+\dot{\hat{r}}) }{\sqrt{\hat{N}^{2} -
 \hat{L}^{2}(\hat{N}^{r}+\dot{\hat{r}})^{2}}}.
\end{align}
The Hamiltonian of the shell is
\begin{align}
\label{22}
H^{s} & = \sqrt{(\hat{\pi}/\hat{L})^{2} + m^{2}}\delta(r- \hat{r}),
\end{align}
and its momentum  is
\begin{align}
\label{23}
H_{r}^{s} & = \hat{\pi} \delta(r- \hat{r}).
\end{align}
Inside and outside of the shell, the constraints are the same as in vacuum.
On the shell, there is a singular contribution to the gravity part of the constraints and it has to be combined with the shell contribution. As a result, we obtain the shell constraints which are
\begin{align}
\label{24}
C^{s} & = \frac{[R']}{L}+\sqrt{(\hat{\pi}/\hat{L})^{2} + M^{2}},
\end{align}
and
\begin{align}
\label{25}
C_{r}^{s} & =L [P_L] +\hat{\pi},
\end{align}
where the jump of a field across the shell is denoted by square  brackets. The next step is to solve the constraints for the inner and outer regions and substitute the solution back into the action. This can easily be done using the Kucha\v{r} variables.
The result is that the bulk terms cancel out as well as the shell term, and all that remains is the  boundary terms that appear as a result of the Kucha\v{r} canonical transformation.
\begin{align}
\label{26}
S & =   \int dt \Big[ m \dot {\hat T} + [P_{R}] \dot{R} -N^s C^s \Big].
\end{align}
Here, $ {\hat T}$ is  the value of the Killing time at the shell,
\begin{align}
\label{27}
{P_R}\Big\vert_{in,out}=\ln \Bigg| \frac{R' + L P_{L}}{R' - L P_{L}}  \Bigg|_{in,out},
\end{align}
and $C^s$ is the constraint (\ref{25}). Then, we reexpress the  constraint (\ref{25}) in terms of the Kucha\v{r} canonical variables $m$ and $P_R$:
\begin{align}
\label{28}
C^s=\sqrt{1-2m+ |\Lambda| R^2}\cosh{P_R}_{out} -\cosh{P_R}_{in}+M,
\end{align}
from which one can find  equations of motion for $R$ as
\begin{align}
\label{29}
\frac{\dot R}{N^s}=\sqrt{1-2m+ |\Lambda| R^2}\sinh{P_R}_{out}=\sinh{P_R}_{in},
\end{align}
which leads to one more constraint
\begin{align}
\label{30}
\sqrt{1-2m+ |\Lambda| R^2}\sinh{P_R}_{out}-\sinh{P_R}_{in}=0.
\end{align}
Substituting (\ref{29}) into (\ref{28}), we rederive the familiar Israel equation:
\begin{equation}\label{31}
\sqrt{1+ |\Lambda| R^2 +\frac{\dot R^2}{(N^s)^2}}+\sqrt{1-2m+ |\Lambda| R^2+\frac{\dot R^2}{(N^s)^2}}-M=0.
\end{equation}
Finally, one can find the single Hamiltonian constraint which describes the dynamics of the shell by taking the sum of squared (\ref{28}) and (\ref{30}):
\begin{align}
\label{32}
1+1-2m-2\sqrt{1-2m+ |\Lambda| R^2}\cosh [P_R]-M^2=0.
\end{align}
It is the constraint that will be used in quantum theory. One can notice that below the horizon, $1-2m+ |\Lambda| R^2<0$, equation (\ref{32}) does not have solutions in real variables. This means that the variables used do not cover the entire phase space. 
Also, the equation contains square roots which means that it is not a single-valued function. Different choices of the signs in front of the square roots correspond to different sectors of the phase space of the model, which are pictured as different regions on the Penrose diagrams.  Besides that, the Killing time $\hat T$ and the radial momentum $[P_R]$ diverge at the horizon.

One way to avoid this problem is to complexify the phase space and to gather different patches into a Riemann surface \cite{vb,vb1}. In the subsequent, we will look for a real chart that would cover the entire phase space of the model.

\section{Global parameterization of ADS and BTZ spacetime}
Both $ADS^3$ spacetime and $BTZ$ spacetime can be given a global parameterization by $SO(2,2)$ group elements, $g$. The metric can be reconstructed from pure gauge $so(2,2)$ connection as
\begin{equation}
A_\mu=g^{-1}\partial_\mu g.
\end{equation}
A connection of $so(2,2)$ is $A_\mu=\Gamma_{AB}A^{AB}_\mu$,  where $\Gamma_{AB}$ are $so(2,2)$ generators, $A,B=0,1,2,3$ can be decomposed into Lorentz connection $\omega^{ab}_\mu=A^{ab}_\mu$, where $a,b=0,1,2$, and triad $e^a_{\mu}=lA^{3a}_\mu$ where $l=1/\sqrt{\Lambda}$. $SO(2,2)$ generators $\Gamma^{AB}$,$\Gamma^{ab}$ -Lorentz and $\Gamma^{a3}$- translations. Extracting triad from
\begin{equation}
A_\mu=g^{-1}\partial_\mu g,
\end{equation}
where $g\in SO(2,2)$. Line element is defined as
\begin{equation}\label{le1}
ds^2=\eta_{ab}e^a_{\mu}e^b_{\nu}dx^{\mu}dx^{\nu}.
\end{equation}
Alternatively, line element could be obtained from embedding of ADS space into four dimensional flat space with signature $(-,+,+,-)$. Embedding coordinates are defined as $X^A=(gv)^A$, where $v=(0,0,0,l)$. They satisfy $\eta_{AB}X^AX^B=-l^2$, where $\eta_{AB}=diag(-1,1,1-1)$. Line element is then
\begin{equation}\label{le2}
ds^2=\eta_{AB}\partial_{\mu}X^A \partial_{\nu}X^B dx^{\mu}dx^{\nu}.
\end{equation}
Parameterization of spacetime by a group element $g$ is related to the  static coordinate system from the previous section by:
\begin{equation}
g=g_t  g_t  g_\phi  g_R ,
\end{equation}
where in the case of $ADS^3$ spacetime
\begin{align}
g_t &= \cos(t/l)I+\sin(t/l)\Gamma^{03}, \nonumber\\
g_\phi &=\cos(\phi)I+\sin(\phi)\Gamma^{12}, \nonumber\\
g_R &=\sqrt{1+R^2/l^2}I+(R/l)\Gamma^{13},
\end{align}
and
\begin{align}
X^0 &= l\sqrt{1+R^2/(l^2)}\cosh(t/l), \nonumber  \\
X^1 &= l\sqrt{1+R^2/(l^2)}\sinh(t/l), \nonumber\\
X^2 &= R\sin(\phi), \nonumber \\
X^3 &= R\cos(\phi),
\end{align}
and the metric is
\begin{equation}
\label{lest}
ds^2=-\Big(\frac{R^2}{l^2}+1\Big)dt^2+\Big(\frac{R^2}{l^2}+1\Big)^{-1}dR^2+R^2d\phi^2.
\end{equation}
For BTZ solution outside the horizon ($r>\sqrt{2m-1}l$), we have
\begin{align}
\nonumber
g_t &= \cosh(\sqrt{2m-1}t/l)I+\sinh(\sqrt{2m-1}t/l)\Gamma^{01}, \nonumber \\
g_\phi &= \cosh(\sqrt{2m-1}\phi)I+\sinh(\sqrt{2m-1}\phi)\Gamma^{32}, \nonumber \\
g_R &= R/(\sqrt{2m-1}l)I+\sqrt{R^2/((2m-1)l^2)-1}\Gamma^{13},
\label{gout}
\end{align}
and
\begin{align}
X^0 &= l\sqrt{R^2/((2m-1)l^2)-1}\sinh(\sqrt{2m-1}t/l), \nonumber \\ 
X^1 &= l\sqrt{R^2/((2m-1)l^2)-1}\cosh(\sqrt{2m-1}t/l), \nonumber \\
X^2 &= R/\sqrt{2m-1}\sinh(\sqrt{2m-1}\phi), \nonumber \\
X^3 &= R/\sqrt{2m-1}\cosh(\sqrt{2m-1}\phi).
\label{xout}
\end{align}
Substituting this into (\ref{le1}) or (\ref{le2}), one obtains
\begin{equation}
\label{lest1}
ds^2=-\Big(\frac{R^2}{l^2}-2m+1\Big)dt^2+\Big(\frac{R^2}{l^2}-2m+1\Big)^{-1}dR^2+R^2d\phi^2,
\end{equation}
which is the familiar expression for BTZ metric in static coordinates. Continuation of the field $g(x)\in SO(2,2)$ inside the horizon, ($R<\sqrt{2m-1}l$), takes into account the interchange of radial and temporal variables:
\begin{align}
g_t &= \cosh(\sqrt{2m-1}t/l)I+\sinh(\sqrt{2m-1}t/l)\Gamma^{01}, \nonumber \\
g_\phi &= \cosh(\sqrt{2m-1}\phi)I+\sinh(\sqrt{2m-1}\phi)\Gamma^{32}, \nonumber \\
g_R &= R/(\sqrt{2m-1}l)I-\sqrt{1-R^2/((2m-1)l^2)}\Gamma^{03},
\label{gin}
\end{align}
and
\begin{align}
X^0 &= -l\sqrt{1-R^2/((2m-1)l^2)}\cosh(\sqrt{2m-1}t/l), \nonumber  \\ 
X^1 &= -l\sqrt{1-R^2/((2m-1)l^2)}\sinh(\sqrt{2m-1}t/l), \nonumber \\
X^2 &= R/\sqrt{2m-1}\sinh(\sqrt{2m-1}\phi), \nonumber \\  
X^3 &= R/\sqrt{2m-1}\cosh(\sqrt{2m-1}\phi).
\label{xin}
\end{align}
The group fields (\ref{gout}) and (\ref{gin}) are continuously (but not smoothly in terms of $r$ and $t$ variables) glued along the horizon. Substituting  (\ref{gin}) into (\ref{le1}) or (\ref{xin}) into (\ref{le2}) again results in the same BTZ metric (\ref{lest}).

As we shall see in the next section, two of the embedding coordinates $X^a$, $a=0,1$ from (\ref{xout}) and (\ref{xin}) will play the role of canonical coordinates of the shell. The shell radius can be expressed in terms of these coordinates as
\begin{equation}
\label{radius}
R=\sqrt{(2m-1)(X^aX_a+l^2)}.
\end{equation}

\section{Action principle and symplectic form} \label{sec4}
The total action consists of gravity action and shell action.
\begin{equation}
\label{sfull}
S=S_{gr}+S_{shell}.
\end{equation}
Gravity action is the Chern-Simons action for $SO(2,2)$ group. As before $A_\mu=\Gamma_{AB}A^{AB}_\mu$ is $so(2,2)$ connection, and $\langle , \rangle$ is a bilinear form on $so(2,2)$ algebra such that $\langle \Gamma^{AB}, \Gamma^{CD}\rangle=\epsilon^{ABCD}$. Then
\begin{equation}
S_{gr}= \frac{\kappa}{8\pi} \int_M d^3 x \epsilon^{\mu \nu \rho}\langle A_\mu, (\partial_\nu A_\rho +\frac{2}{3} A_\nu A_\rho)  \rangle.
\end{equation}
The shell is divided into $N$ particles with label $i$,
\begin{equation}\label{actionshell}
S_{shell}=\sum\limits_i^N \int_{l_i} Tr(K_i A_\mu) dx^\mu,
\end{equation}
where integration is along each particle worldline $l_i$ and $K_i=m_i \Gamma_{03}$ is an element of $so(2,2)$-algebra. Gravity action is invariant under gauge transformations:
\begin{equation}\label{gaugetrans}
A_\mu \rightarrow g^{-1}(\partial_\mu+A_\mu)g,
\end{equation}
where $g$ is an $SO(2,2)$ group element. Unlike gravity action, shell action is variant under gauge transformations. Each particle is transformed under gauge transformations (\ref{gaugetrans}) as
\begin{equation}\label{wbg}
\int_{l_i} Tr(K_i A_\mu)dx^\mu \rightarrow \int_{l_i} Tr(\tilde K_i A_\mu)dx^\mu+\int_{l_i} Tr( K_i g^{-1}\dot g )d\tau,
\end{equation}
where $\tilde K_i= gK_ig^{-1}$, $\tau$ is a parameter along the particle worldline and the derivative with respect to it is represented by dot. The second term on the r.h.s. of (\ref{wbg}) represents the action  of a massive spinless particle on ADS space. Thus, the  degrees of freedom of the particles are represented by formerly gauge degrees of freedom.

We slice the manifold so that the particle worldlines  are in the direction of time coordinate.  Then the variation of the action (\ref{sfull}) with respect to $A_0$  results in the following constraint
\begin{equation}\label{const1st}
\epsilon^{0\mu \nu}F(A)_{\mu \nu}=\sum\limits_i^N \tilde K_i \delta^2 (x,x_i),
\end{equation}
where $F(A)_{\mu \nu}$ is the curvature of connection $A$, and $x_i$ is the location of each particle. We have to choose one component of connection $A$  to be zero to linearize the constraint.  Such  choice of a gauge cannot be globally made, because the model contains a non-trivial moduli space.

Following \cite{amms,amms1}, the  spacial slice is divided into different regions in which such choice of a gauge could be made. Each region is surrounded by a circle containing only one particle. Note that there is no common boundaries between the circles and they are connected to common origins, as shown in Fig.1. By making cuts along the circles, the manifold is splitted into $N$ discs and a polygon. Each disc contains a particle, while a polygon contains no particles, but connected to infinity.
\begin{figure}
\includegraphics[scale=0.65]{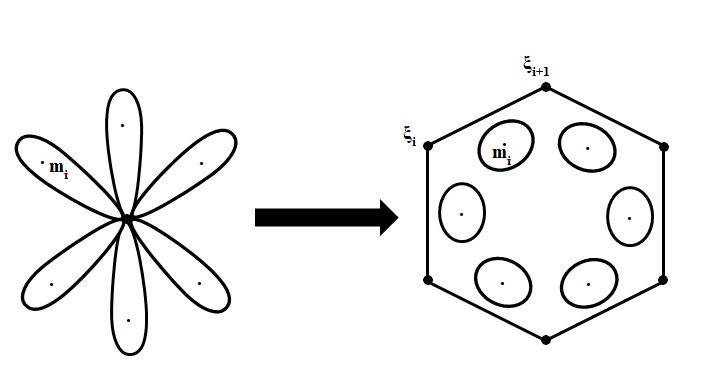}
\centering
\caption{Formation of discs and a polygon due to splitting of space.}
\end{figure}

The solution for the discs is written in polar coordinates. Then, we select the gauge in which the radial component of $A$ has to be zero. Next, we solve the constraints and plug the solution back into the action in an arbitrary gauge:
\begin{equation}\label{constsol}
A_{r,i}=g_i^{-1}\partial_r g_i \ \ \ \ A_{\phi,i}=g_i^{-1}\nabla_\phi g_i,
\end{equation}
where $\nabla_\phi g_i=\partial_\phi g_i+K_i g_i$. Similarly, for the polygon, in which the gauge parameter will be labeled as $h$. After solving the constraints all that remains of the action is the kinetic term:
\begin{equation}\label{kinterm}
S_{D_i}=\int_{D_i} d^3x \epsilon^{0\mu \nu} \langle A_\mu \dot A_\nu \rangle + \int_{l_i} Tr( K_i  g_i^{-1} \dot g_i)d\tau,
\end{equation}
and similar for the polygon, without a source. However, it is much easier to calculate the symplectic form which is the variation of the kinetic term of the action:
\begin{equation}\label{sympf}
\Omega_{D_i}=\int_{D_i} d^2x \epsilon^{0\mu \nu} \langle \delta A_\mu \wedge \delta A_\nu \rangle + \delta Tr( K_i \delta g_i^{-1} \wedge \delta g_i).
\end{equation}
After plugging in the constraint solution in the form (\ref{constsol}) and using the identity $\delta (g_i^{-1}\nabla_\mu g_i)=g_i^{-1}\nabla_\mu(\delta g_i g_i^{-1}) g_i$, the symplectic form for the disk reduces to its boundary:
\begin{equation}\label{sympfd}
\Omega_{ D_i}=\int_{\partial D_i} dx  \langle\nabla_\phi ( \delta g_i g_i^{-1}). \wedge \delta g_i g_i^{-1}\rangle.
\end{equation}
The same situation for polygon, whose symplectic is a sum of contributions from every edge $E_i$:
\begin{equation}\label{sympfp}
\Omega_{P}=\sum\limits_i^N \int_{E_i} dx  \langle\partial_\phi ( \delta h_i h_i^{-1}) \wedge \delta h_i h_i^{-1}\rangle.
\end{equation}
Next, we have to collect all the symplectic form together and consider the condition of continuity of metric and connection between discs and polygon. Firstly, covariant derivative in (\ref{constsol}) has to be converted to ordinary derivative using gauge transformation
\begin{equation}\nonumber
\tilde g_i = \exp(K\phi)  g_i.
\end{equation}
The symplectic form of disc (\ref{sympfd}) changes to
\begin{equation}\label{kintermbt}
\Omega_{ D_i}=\int_{\partial D_i} dx   \langle\nabla_\phi ( \delta \tilde g_i \tilde g_i^{-1}) \wedge \delta \tilde g_i \tilde g_i^{-1}\rangle.
\end{equation}
The continuity conditions for the connection  (\ref{constsol}) is
\begin{equation}\label{overlap}
\tilde g_i=C_i h\Big\vert_{E_i},
\end{equation}
where $C_i$ is a  function only of time.  Substituting this into (\ref{sympfd}) and (\ref{sympfp}), and combining them one obtains
\begin{equation} \label{sympfful}
\Omega_{full}=\Omega_P+\sum\limits_i^N \Omega_{D_i}= \sum\limits_i^N \int_{E_i}\langle\partial_\phi( \delta h h^{-1}), C_i^{-1}\delta C_i\rangle=-
\sum\limits_i^N \int_{\partial D_i}\langle\partial_\phi( \delta \tilde g_i \tilde g_i^{-1}), \delta C_i C_i^{-1}\rangle.
\end{equation}
This symplectic form collapses to the vertices of the polygon or to the initial points of disc boundaries.
Now, let us rewrite the action in the following  variables 
\begin{equation}\label{nvar}
u_i=C_i^{-1}\exp(2\pi K)C_i, \ \ {\rm and} \ \ \ \ \ \  h_i = C_i^{-1} \tilde g_i (0),
\end{equation}
where
\begin{equation}\nonumber
\tilde g_i (2\pi) = \exp(2\pi K)\tilde g_i (0),
\end{equation}
and
\begin{equation}\nonumber
u_i^{-1}\delta u_i = C_i^{-1}\delta C_i - C_i^{-1}\exp(-2\pi K)\delta C_i C_i^{-1} \exp(2\pi K) C_i.
\end{equation}
Equation (\ref{sympfful}) reduces to
\begin{eqnarray}\label{sympffull1}
\Omega_{full}= \sum\limits_i^N \int_{R}\langle \delta h_{i} h_{i}^{-1} , u_i^{-1}\delta u_i\rangle.
\end{eqnarray}
$h_i$ will play the role of configuration variable and $u_i$-momentum variable. The translational part of $h_i$ gives rise to the BTZ metric outside of the shell. The canonical coordinates are thus the variables (\ref{xout}) or (\ref{xin}). By using  the overlap conditions (\ref{overlap}) which implies that
\begin{equation}\nonumber
g_i (0) = C_i h_i, \ \ \ g_i (2\pi)=\exp(2\pi K) g_i (0)=C_i h_{i+1},
\end{equation}
we find that
\begin{equation}
g_{i+1}=u_i g_i ,
\end{equation}
and
\begin{equation}\label{xii0}
g_{i}=\left( \prod\limits_{j=0}^{i-1} u_j\right) g_0 \left( \prod\limits_{j=0}^{i-1} u_j \right)^{-1},
\end{equation}
where the order of factors in the product are from right to left. The holonomy around the full shell is defined as the product of holonomies around each particle
\begin{equation}\label{ufull}
 U=\prod\limits_{j=0}^{N} u_i.
\end{equation}
By using the following identity
\begin{equation}\nonumber
U^{-1}\dot U =\sum\limits_{i=0}^N \left( \prod\limits_{j=0}^{i-1} u_j \right)^{-1} u_i^{-1}\dot u_i\left( \prod\limits_{j=0}^{i-1} u_j\right),
\end{equation}
we can rewrite the symplectic form (\ref{sympffull1}) as
\begin{equation}\label{sympffull}
\Omega_{full}=
 \int_{R }\langle \delta h_0 h_0^{-1},   U^{-1}\delta  U \rangle.
\end{equation}
The symplectic form for full shell is reduced to a term depending on a single variable. It can be shown that only translational part of $h$ and the only Lorentzian part of $U$ will enter (\ref{sympffull}). In the neighborhood of the horizon where $\delta h_0 h_0^{-1} \ll 1$  the symplectic form simplifies further:
\begin{equation}\label{sympffulla}
\Omega_{full}=
\frac{1}{l} \int_{R }\langle \delta X ,   U^{-1}\delta  U \rangle,
\end{equation}
where $X=X^a\Gamma_a$ and $X_a$ are embedding coordinates  (\ref{xout}) or (\ref{xin}) for $\phi=0$.

\section{Constraints} \label{sec5}
Here, we find the equations to which the holonomy around the full shell $U$ is satisfied. Let $C_i$ and $u_i$ are elements of Anti-de sitter group $SO(2,2)$. By the definition of $U$ which is the product of holonomies around every particle, we have
\begin{align}
\label{const2}
U & = \prod\limits_{i=0}^{N}  u_i,
\end{align}
where
\begin{align}
\label{const3}
u_i & =C_{i}^{-1} \exp\bigg( \frac{2 \pi M}{N} \Gamma^{12} \bigg) C_{i},
\end{align}
which represents the holonomy around a fixed particle. Choose the following ansatz for $C_{i}$, 
\begin{align}
\label{const4}
C_{i} & = (tr)_i (b)_i,
\end{align}
where $tr$ and $b$ refer to translations and boost parameters.
\begin{align}
\label{const5}
(tr)_i & =  \exp\bigg( \frac{-2 \pi i}{N} \Gamma_{12} \bigg) (tr)_{0}  \exp\bigg( \frac{2 \pi i}{N} \Gamma_{12} \bigg),
\end{align}
\begin{align}
\label{const6}
(b)_i & =  \exp\bigg( \frac{-2 \pi i}{N} \Gamma_{12} \bigg) (b)_{0}  \exp\bigg( \frac{2 \pi i}{N} \Gamma_{12} \bigg),
\end{align}
\begin{align}
\label{const7}
(tr)_{0} & =  t  = \sqrt{1+\Lambda R^2}  I +  \sqrt{\Lambda} R  \Gamma^{13},
\end{align}
\begin{align}
\label{const8}
(b)_{0} & = b  =  \cosh \bar{x} I + \sinh \bar{x}  \Gamma^{10}.
\end{align}
Then, the product of holonomies of two neighboring particles is
\begin{align}
\label{const9}
u_i u_{i+1} & = C_{i}^{-1} K C_{i} C_{i+1}^{-1} K C_{i+1},
\end{align}
where
\begin{align}
\label{const10}
K & =  \exp\bigg( \frac{2 \pi M}{N} \Gamma^{12} \bigg).
\end{align}
Using equations (\ref{const3}), (\ref{const4}), (\ref{const5}), (\ref{const6}), (\ref{const7}) and  (\ref{const8}), equation (\ref{const9}) can be simplified as
\begin{align}
\label{const11}
u_i u_{i+1} & =  \exp\bigg( \frac{-2 \pi i}{N} \Gamma_{12} \bigg) (b)^{-1} (t)^{-1} K t b  \exp\bigg( \frac{2 \pi i}{N} \Gamma_{12} \bigg) \nonumber \\ & \exp\bigg( \frac{-2 \pi (i+1)}{N} \Gamma_{12} \bigg) (b)^{-1} (t)^{-1} K t b  \exp\bigg( \frac{2 \pi (i+1)}{N} \Gamma_{12} \bigg).
\end{align}
In terms of equation (\ref{const11}), the holonomy around the full shell (\ref{const2}) can be written as
\begin{align}
\label{const12}
U & = \prod\limits_{i=0}^{N}  u_i  = \prod\limits_{i=0}^{N} \exp\bigg( \frac{-2 \pi i}{N} \Gamma_{12} \bigg) (b)^{-1} (t)^{-1} K t b, \nonumber \\ &=  \Bigg[ \exp\bigg( \frac{-2 \pi i}{N} \Gamma_{12} \bigg) (b)^{-1} (t)^{-1} K t b \Bigg]^{N}.
\end{align}
Note that
\begin{align}
\label{const13}
(t)^{-1} K t & =  \exp\bigg[ \frac{M}{N} \Big( \sqrt{1+|\Lambda| R^2}  \Gamma_{12}^{-1} +  \sqrt{|\Lambda|} R  \Gamma_{23} \Big) \bigg], \\
\label{const14}
 (b)^{-1} (t)^{-1} K t b & = \exp\bigg[ \frac{M}{N} \Big( \sqrt{1+|\Lambda| R^2} \cosh \bar{x} \Gamma_{12} + \sqrt{1+|\Lambda| R^2} \sinh \bar{x} \Gamma_{02} +  \sqrt{|\Lambda|} R  \Gamma_{23} \Big) \bigg].
\end{align}
Then, equation (\ref{const12}) is
\begin{align}
\label{const15}
U & =  \exp\bigg[ 2 \pi \Big( 1- M \sqrt{1+|\Lambda| R^2} \cosh \bar{x}  \Big) \Gamma_{12} + M \sqrt{1+|\Lambda| R^2} \sinh \bar{x} \Gamma_{02} +  M \sqrt{|\Lambda|} R  \Gamma_{23}  \bigg].
\end{align}
Then, we take the trace of equation (\ref{const15}), we find that
\begin{align}
\label{const16}
Tr(U) & =  \cos \bigg[ 2 \pi \sqrt{ \Big( 1- M\sqrt{1+|\Lambda| R^2} \cosh \bar{x}  \Big)^{2} - \Big( M \sqrt{1+|\Lambda| R^2} \sinh \bar{x} \Big)^{2} - \Big( M \sqrt{|\Lambda|} R  \Big)^{2} } \bigg].
\end{align}
In terms of the ADM variables, we have
\begin{align}
\label{const17}
Tr(U) & =  \cos \big[ 2 \pi  \sqrt{1 - 2m} \big].
\end{align}
Compare (\ref{const16}) with (\ref{const17}), we have
\begin{align}
\label{const18}
1- 2m & =   \Big( 1- M\sqrt{1+|\Lambda| R^2} \cosh \bar{x}  \Big)^{2} - \Big( M \sqrt{1+|\Lambda| R^2} \sinh \bar{x} \Big)^{2} - \Big( M^{2} |\Lambda| R^{2}  \Big)^{2}.
\end{align}
We find that
\begin{align}
\label{const19}
m & =  M \Big( \sqrt{1+|\Lambda|R^2} \cosh \bar{x} - \frac{M}{2}   \Big),
\end{align}
and
\begin{align}
\label{const20}
M & =  \sqrt{1+ |\Lambda| R^2} \cosh \bar{x}  \pm  \sqrt{(1+ |\Lambda| R^2) \cosh^{2} \bar{x} - 2m+1}.
\end{align}
Notice that $R$ in the transformation (\ref{const7}) was used as a longitudinal radial variable. This variable undergoes the Lorentz contraction under radial boost transformations. On the other hand, the perimeterial radius entering (\ref{13})  is transverse to the radial boost and is thus invariant. To take this into account, we have to undergo the Lorentz contraction by rescaling $R \rightarrow R \cosh \bar{x}$. Then, in terms of the perimeterial radius, equation (\ref{const20}) becomes
\begin{align}
\label{const21}
M & =  \sqrt{1+|\Lambda| R^2 + \sinh^{2} \bar{x}}   \pm  \sqrt{1 - 2m +|\Lambda| R^2 + \sinh^{2} \bar{x}},
\end{align}
which is the familiar Israel equation.

\section{Derivation of the constraint equations} \label{sec6}
In the previous section, we obtained the expression for the holonomy around the shell. To obtain the Hamiltonian constraint in the canonical form, we have to extract the Lorentzian part of the above holonomy and then reexpress it in terms of the Euler angles which provide global parameterization of the Lorentzian manifold.

To extract the Lorentzian part of the holonomy, one has to perform its conjugation by a translation transformation
\begin{align}
\label{conj}
U_l=g_t^{-1}Ug_t,
\end{align}
where
\begin{align}
\label{ualg}
U=\exp(\pi J), \ \ J=(1-M\sqrt{1+\frac{R^2}{l^2}}\cosh(\bar \chi))\Gamma_{12}+\sinh(\bar\chi)\Gamma_{20}+M\cosh(\bar\chi)\frac{R}{l}\Gamma_{23}.
\end{align}
Note that $g_t$ in (\ref{conj}) is to be found from the condition that $g_t^{-1}Jg_t$ is a  pure Lorentz transformation. This results in an equation for $g_t$ which could be sought for in a form $g_t=aI+\sqrt{-1+a^2}\Gamma_{13}$. Plugging this into (\ref{ualg}), one finds $a=R/(\sqrt{2m-1}l)$, or
\begin{align}
g_t=R/(\sqrt{2m-1}l)I+\sqrt{R^2/((2m-1)l^2)-1}\Gamma^{13}=\sqrt{X^2/l+1}I+X/l\Gamma^{13},
\end{align}
which coincides with the radial translation in BTZ space (\ref{xout}), $X^2=X_aX^a$. Now, one has to express the total mass $m$ in terms of canonical coordinates $X^a$.
From
\begin{align}
m  =  M \Big( \sqrt{\cosh^2 \bar{\chi}+\frac{ R^2}{l^2}}  - \frac{M}{2}   \Big),
\end{align}
and
\begin{equation}
R^2=(2m-1)(X^aX_a+l^2),
\end{equation}
one obtains the following equation for $m$:
\begin{align}
m  =  M \Big( \sqrt{2m(1+\frac{X^2}{l^2}) + \sinh^2 \bar{\chi}-\frac{X^2}{l^2}}  - \frac{M}{2}   \Big).
\end{align}
We shall solve this equation in the limit of the slow movement, $\bar \chi \ll 1$, and near-horizon location $X^2 \ll l^2$
where it becomes
\begin{align}
m  =  M \Big( \sqrt{2m}+\frac{2}{\sqrt{2m}}((2m-1)\frac{X^2}{l^2} + \sinh^2 \bar{\chi})  - \frac{M}{2}   \Big).
\end{align}
The approximate solution is
\begin{align}
m  =  M/\sqrt{2}   + \frac{\sqrt{2}}{\sqrt{M}}((\sqrt{2}M-1)\frac{X^2}{l^2}+\sinh^2 \bar{\chi})).
\end{align}
Next, we can express $R$ as
\begin{equation}
R=\sqrt{M/\sqrt{2}-1}l +a(M)X^2/l +b(M)\bar{\chi}^2 l,
\end{equation}
where $a(M)$ and $b(M)$ are constants related to the bare mass. Now, we can write an expression for $U_l$ in terms of canonical coordinates and boost parameter
\begin{equation}
U_l =\exp(\pi J_l),
\end{equation}
where
\begin{equation}
 J=\Big((1-M\sqrt{\cosh^2(\bar \chi)+\frac{R^2}{l^2}})\sqrt{X^2/l^2+1}+M\frac{RX}{l^2}\Big)\Gamma_{12}+M\sinh(\bar\chi)\Gamma_{20}.
\end{equation}
In terms of the Euler angles $\rho$ and $\chi$
\begin{align}
\cos(\rho)\cosh(\chi) &= 
\cos(\pi\sqrt{\Big((1-M\sqrt{\cosh^2(\bar \chi)+\frac{R^2}{l^2}})\sqrt{X^2/l^2+1}+M\frac{RX}{l^2}\Big)^2-\Big(M\sinh(\bar\chi)\Big)^2}),
\end{align}
and
\begin{align}
\sinh(\chi) &= 
\frac{\sinh{\bar \chi}\sin(\pi\sqrt{\Big((1-M\sqrt{\cosh^2(\bar \chi)+\frac{R^2}{l^2}})\sqrt{X^2/l^2+1}+M\frac{RX}{l^2}\Big)^2-\Big(M\sinh(\bar\chi)\Big)^2})}
{\sqrt{\Big((1-M\sqrt{\cosh^2(\bar \chi)+\frac{R^2}{l^2}})\sqrt{X^2/l^2+1}+M\frac{RX}{l^2}\Big)^2-\Big(M\sinh(\bar\chi)\Big)^2}}.
\end{align}
From the second expression we express $\bar \chi$ in terms of canonical variables in the limit $\bar \chi \ll 1$
\begin{align}
\sinh(\bar \chi) &= 
\frac{\sinh(\chi)\Big((1-M\sqrt{1+\frac{R^2}{l^2}})\sqrt{X^2/l^2+1}+M\frac{RX}{l^2}\Big)}
{\sin(\pi\Big((1-M\sqrt{1+\frac{R^2}{l^2}})\sqrt{X^2/l^2+1}+M\frac{RX}{l^2}\Big))}.
\end{align}
After all substitutions the Hamiltonian constraint will take the form
\begin{eqnarray}
\label{hfin}
\cos(\rho)=\frac{\cos(\pi\sqrt{a_1(M)-a_2(M)\chi^2+a_3(M)X^2/l^2})}{\cosh(\chi)},
\end{eqnarray}
where $a_1$, $a_2$, $a_3$ are coefficients depending on $M$ only.

\section{Quantization} \label{sec7}
We shall perform quantization in a neighborhood of the horizon $X_0,X_1 \ll l$ in (\ref{xout}) or (\ref{xin}), where the translation non-commutativity can be neglected and a standard momentum representation could be constructed.
\begin{equation}
U=uI +p^a\gamma_a,
\end{equation}
where
\begin{equation}
u=\cosh\chi \cos \rho, p_0=\cosh\chi \sin \rho,p_1=\sinh\chi \cos \phi,p_2=\sinh\chi \sin \phi.
\end{equation}
Coordinate operators are
\begin{align}
\hat x^a U &= U\gamma_a, \\
\gamma_a\gamma_b &=\eta_{ab}I +\epsilon_{abc}\gamma^c,
\end{align}
from which we can deduce
\begin{align} \label{lir}
\hat X^0 \Psi &=i(\frac{\partial}{\partial \rho}-\frac{\partial}{\partial \phi}) \Psi, \nonumber \\
\hat X^1 \Psi &=i(\cos\rho\cos\phi\frac{\partial}{\partial \chi}
+\tanh\chi\sin\rho\cos\phi\frac{\partial}{\partial \rho}
+\coth\chi\cos\rho\sin\phi\frac{\partial}{\partial \phi}) \Psi, \nonumber \\
\hat X^2 \Psi &=i(\cos\rho\sin\phi\frac{\partial}{\partial \chi}
+\tanh\chi\sin\rho\sin\phi\frac{\partial}{\partial \rho}
-\coth\chi\cos\rho\cos\phi\frac{\partial}{\partial \phi}) \Psi.
\end{align}
By using a cylindrical symmetry ($\frac{\partial}{\partial \phi}=0$, and only two components of $X$ remains)
\begin{align}
\hat X^0 \Psi & =i(\frac{\partial}{\partial \rho}) \Psi, \\
\hat X^1 \Psi &=i(\cos\rho\frac{\partial}{\partial \chi}
+\tanh\chi\sin\rho\frac{\partial}{\partial \rho}) \Psi.
\end{align}
The kinematical states of the model is defined as functions of $U\in ADS^2$, 
\begin{equation}
\Psi(U)=\Psi(\rho,\chi),
\end{equation}
which is a single-valued and periodic in $\rho$ functions on the entire momentum space. The scalar product can be deduced from the Haar measure on $SL(2)$ and is defined as
\begin{equation}
\langle\Phi, \Psi \rangle= \frac{1}{\pi}\int \cosh(\chi)d \rho d\chi \Phi(\rho,\chi)^* \Psi(\rho,\chi).
\end{equation}
The spectrum of time coordinate $X^0$ is canonically conjugate to $\rho$, and its corresponding operator is
\begin{equation}
\hat T |\rho,\chi\rangle=i\hbar\frac{\partial}{\partial\rho} |\rho,\chi\rangle,
\end{equation}
and its eigenstates are
\begin{equation}
|t;\psi\rangle=\frac{1}{\pi}\int \cosh(\chi)d \rho d\chi \exp(it\rho)\psi(\chi)|\rho,\chi\rangle,
\end{equation}
where $t$ is an integer. Time operator has only a discrete spectrum:
\begin{equation}
\hat T|t;\psi\rangle=t\hbar  |t;\psi\rangle.
\end{equation}
Notice that Newton constant $G$ is equal to one and the quantization is in the units of the Planck length.

A more interesting observable is a Lorentz-invariant length, $X^2= X_a X^a$,  defining the distance of the shell to the horizon. This distance is related to the perimeterial radius of the shell by (\ref{radius}). Notice that the peremeterial radius depends not only on canonical coordinates $X^a$ but also on the total mass $m$ which is a function of canonical momenta. Because of this $R$ is not diagonalizable simultaneously with $X$. We shall choose the basis in which $X$ is diagonal as more convenient.

As $X_a$ in (\ref{lir}) is specified as the left-invariant derivative on the group and its square is represented by the Beltrami-Laplace operator on our momentum space:
\begin{equation}
\hat X^2 |t;\psi\rangle=  |t;\Delta \psi\rangle,
\end{equation}
where
\begin{equation}
\label{bellap}
\Delta=\hbar^2\left(\frac{1}{\cosh(\chi)}\frac{\partial}{\partial\chi}\cosh(\chi)\frac{\partial}{\partial\chi}
+\frac{t^2}{\cosh^2(\chi)} \right).
\end{equation}
This operator has two series of eigenvalues as shown in \cite{mw}. One is positive, but separated from zero, corresponds to continuous spectrum,i.e. spacelike, $X^2$
\begin{equation}
\hat X^2 |t,\lambda \rangle=\pi(\lambda^2+1/4)\hbar^2 |t,\lambda \rangle,
\end{equation}
where $\lambda$ is a real number. The other is negative, but containing zero, corresponds to discrete spectrum, i.e. timelike, $X^2$
\begin{equation}
\hat X^2 |t,l \rangle=-\pi l(l+1)\hbar^2 |t,l \rangle,
\end{equation}
where $l$ is a non-negative integer, subject to the condition $l\leq t$. As it is seen from (\ref{radius}) that positive $X^2$ corresponds to the shell outside the horizon, while negative $X^2$ corresponds to the shell inside the horizon. Thus, the shell radius takes on a continuous set of values outside the horizon and discrete inside.

\section{Quantum dynamics} \label{sec8}
Angular variable $\rho$ in (\ref{hfin}) plays the role of Hamiltonian canonically conjugate to time-variable $t$. In coordinate representation, it become a time derivative operator $\rho \rightarrow i\partial /(\partial t)$. Plugging this into (\ref{hfin}), we see that quantum Hamiltonian constraint is not a differential equation but a finite difference equation 
\begin{eqnarray}
\label{hfinq}
\frac{1}{2}(\Psi(t+1,\chi)+\Psi(t+1,\chi))=\frac{\cos(\pi\sqrt{a_1(M)-a_2(M)\chi^2+a_3(M)\Delta/l^2})}{\cosh(\chi)}\Psi(t,\chi),
\end{eqnarray}
where we use a skew representation (coordinate in time variable and momentum in a spatial variable) and $\Delta$ is the Beltrami-Laplace operator in momentum space. This is a discrete analog of the Klein-Gordon equation which is reduced to the ordinary differential Klein-Gordon equation in zero gravity limit. Concerning the factor ordering issue in the above expression we always choose symmetric order in to render the Hamiltonian hermitian and the evolution operator unitary.

The discrete analog of the Schrodinger equation can be written by using the evolution operator for one step in time 
\begin{equation}
\Psi(t+1,\chi)= U^{\pm} \Psi(t,\chi),
\end{equation} 
where $U$ was found from (\ref{hfinq}) to be
\begin{equation}
U^{\pm}=F\pm\sqrt{F^2-1},
\end{equation} 
where
\begin{equation}
\label{fdef}
F=\frac{\cos(\pi\sqrt{a1(M)-a2(M)\chi^2+a3(M)\Delta/l^2})}{\cosh(\chi)}.
\end{equation} 

\begin{figure}
\label{pendiag}
\includegraphics[scale=0.65]{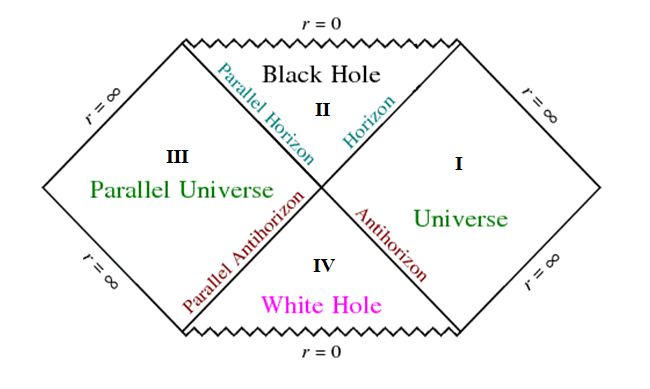}
\centering
\caption{Different four regions of Penrose diagram}
\end{figure}

Now, we are ready to calculate transition amplitudes between different locations of the shell in spacetime. As we saw in the previous section, the location of the shell could be described by two quantum numbers: time coordinate $t$ and the eigenvalue of the invariant distance to the horizon $X^2=X_a X^a$. The corresponding state in the momentum representation is
\begin{equation}
\label{kstate}
|t, X^2 \rangle= \exp(i t \rho) L_{t,X^2}(\chi) |\rho, \chi \rangle,
\end{equation}
where $L_{t,X^2}(\chi)$ is the eigenstate of the operator (\ref{bellap}) with the eigenvalue $X^2$. Now, we can convert the kinematical state (\ref{kstate}) into a physical state by applying the Hamiltonian constraint (\ref{hfin})
\begin{equation}
\label{kstate1}
|t, X^2 \rangle_{phys}= (a_+ (U^+)^t+ a_-(U^-)^t) L_{t,X^2}(\chi) |\rho, \chi \rangle,
\end{equation}
where $a_+$ and $a_-$ are arbitrary coefficients.

One can distinguish four types of such states.
The states with $X^2>0$, $a_+ \neq 0$, $a_-=0$ correspond to sector $I$ of the Penrose diagram in Fig. 2. In zero gravity limit, this sector corresponds to positive frequency solutions of the Klein-Gordon equation. The states with $X^2>0$, $a_+ = 0$, $a_-\neq 0$ correspond to sector $III$ in Fig.2. In zero gravity limit, this sector corresponds to negative frequency solutions of the Klein-Gordon equation. The states with $X^2<0$, $a_+ \neq 0$, $a_-=0$ correspond to sector $II$ in Fig.2. The zero gravity limit keeps no trace of such kind of states. The states with $X^2<0$, $a_+ = 0$, $a_-\neq 0$ correspond to sector $IV$ in Fig.2. The zero gravity limit keeps no trace of such kind of states. We calculate numerically matrix elements between various states of the type
\begin{equation}
\langle t_1 X^2_1|t_2, X^2_2 \rangle_{phys},
\end{equation}
which describe the rate of change of the sell radius from $X^2_2$ to $X^2_1$ during the time interval $t_1-t_2$.
\begin{figure}
\label{result1}
\includegraphics[scale=0.65]{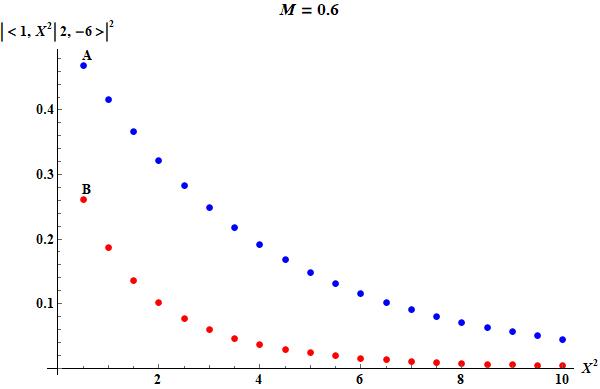}
\centering
\caption{$II \rightarrow I$ (curve B) vs. $I \rightarrow II$ (curve A) transition rate}
\end{figure}

For example, in Fig.3, we show the relative rate for the shell to cross the horizon from region $I$ to region $II$ and back. As one can see that the transition rate $II \rightarrow I$ is comparable to that of  $I \rightarrow II$ 
in a close vicinity of the horizon, but become exponentially damped away from the horizon. This agrees with the results obtained earlier in \cite{vb} by a very different method.

\section{Conclusion}
Quantum theory of a model describing a dust shell coupled to 2+1- dimensional gravity with negative cosmological constant has been studied both at the kinematical and dynamical level in a near-horizon region.
 
 At the kinematical level, it was shown that the shell radius has a continuous spectrum outside the horizon and discrete inside. The eigenvalues spacing of the shell radius measured along the radial coordinate $X$ is Planckian, while for the radius measured along the perimeter of the shell the eigenvalue spacing is proportional to the square root of the black hole mass. 
  
Although, the approximation used does not allow us to go deep inside the black hole, 
one can tell that the point of the central singularity, the zero radius of the shell, belongs to the discrete spectrum. This is suggestive for the singularity resolution.
   
At the dynamical level, we obtained transition amplitudes between different locations of the shell in the near-horizon region. It was shown that there is a non-zero transition rate between all possible sectors of the Penrose diagram, even between those which are classically forbidden. However for the classically forbidden transitions their rate is exponentially damped away from the horizon.

The main reason for studying the cylindrically symmetric shell model is that it could possibly be extended to 3+1 spacetime dimensions. Some results exist on quantum kinematics of a Schwarzschild black hole in a frame of a test particle \cite{drps,drps1}. It has also such features like coordinate non-commutativity and discreteness. Even though the many body problem in 3+1 gravity is not solvable, the holonomy composition  still could be  as in sections 5,6. The only difference is the presence of Newtonian potential. Thus, in principle, there is a possibility to generalize the above results to 3+1 dimensional gravity.

\end{document}